\documentclass[12pt]{article}
\begin{document}

\title{\bf Teleparallel Killing Vectors of the Einstein Universe}

\author{M. Sharif \thanks{msharif@math.pu.edu.pk} and M. Jamil Amir
\thanks{mjamil.dgk@gmail.com}\\
Department of Mathematics, University of the Punjab,\\
Quaid-e-Azam Campus, Lahore-54590, Pakistan.}

\date{}

\maketitle

\begin{abstract}
In this short paper we establish the definition of the Lie
derivative of a second rank tensor in the context of teleparallel
theory of gravity and also extend it for a general tensor of rank
$p+q$. This definition is then used to find Killing vectors of the
Einstein universe. It turns out that Killing vectors of the Einstein
universe in the teleparallel theory are the same as in General
Relativity.
\end{abstract}

{\bf Keywords:} Teleparallel Theory, Killing Vectors.

\section{Introduction}

In General Relativity (GR), the importance of symmetry is quite
clear. The symmetry restrictions are very much helpful to find the
solution of the Einstein field equations (EFEs). Much attention
has been given to study the different kinds of symmetries during
the last three decades. As a pioneer, Petrov [1] solved the
Killing equations for four-dimensional spaces. Bokhari and Qadir
[2-3] were able to achieve a complete classification of static
spherically symmetric spacetimes. Later, Qadir and Ziad extended
this work for a complete classification of all spherically
symmetric spacetimes [4-5] by removing the condition of staticity.
The solutions of the EFEs, corresponding to different symmetries
possessed by the metric tensor, have further been classified
according to their properties and groups of motion admitted by
them [6].

Katzin et al. [7-8] studied the curvature and Ricci collineations
(RCs) in the context of the related particle and field
conservation laws. A detail investigation of the spacetimes and
their geometrical symmetries like Killing vectors (KVs), curvature
collineations and RCs was made by different authors [9-13]. Using
Lie algebra approach, Carot et al. [14] discussed the physical
properties of the spacetimes called matter collineations (MCs).
Camci and Sharif [15] worked out the MCs of homogeneous
G$\ddot{o}$del-type metrics. One of the authors (MS) [16-17] found
the MCs of the spherically symmetric spacetimes (both static and
non-static). Later, this work was extended to classify static
plane and cylindrically symmetric spacetimes according to their
MCs by the same author [18-19].

Teleparallel theory of gravity (TPG) is an alternative theory of
gravity which corresponds to a gauge theory of translation group
[20] based on Weitzenb$\ddot{o}$ck geometry [21]. This theory is
characterized by the vanishing of curvature identically while the
torsion is taken to be non-zero. In TPG, gravitation is attributed
to torsion which plays a role of force [22-25]. In GR, gravitation
geometrizes the underlying spacetime. The translational gauge
potentials appear as a non-trivial part of the tetrad field and
induce a teleparallel (TP) structure on spacetime which is
directly related to the presence of a gravitational field. In some
other theories [22-27], torsion is only relevant when spins are
important. This point of view indicates that torsion might
represent additional degrees of freedom as compared to curvature

There is a large body of literature available [28-30 and
references there in] about the study of TP versions of the exact
solutions of GR. Pereira, et al. [30] obtained the TP versions of
the Schwarzschild and the stationary axisymmetric Kerr solutions
of GR. They proved that the axial-vector torsion plays the role of
the gravitomagnetic component of the gravitational field in the
case of slow rotation and weak field approximations. In recent
papers [31-33], we have found the TP versions of the Friedmann
models, Lewis-Papapetrou spacetimes and stationary axisymmetric
solutions. The energy-momentum distribution of these versions have
also been discussed.

This paper is devoted to look at the symmetry of a metric tensor in
the context of TPG. For this purpose, we define the TP version of
the Lie derivative which gives Killing equations in TPG. This has
been used to find the TP KVs of the Einstein universe giving the
comparison of KVs with GR. The paper is organized as follows.
Section $2$ contains an overview of the TP theory. In section $3$,
we define the Lie derivative in the framework of TPG and the Killing
equations. Section $4$ is devoted to explore the TP KVs of the
Einstein universe. The last section concludes the results obtained.

\section{An Overview of the Teleparallel Theory}

The TP theory is based on the Weitzenb$\ddot{o}$ck connection
given as [25]
\begin{eqnarray}
{\Gamma^\theta}_{\mu\nu}={{h_a}^\theta}\partial_\nu{h^a}_\mu,
\end{eqnarray}
where ${h_a}^\nu$ is a non-trivial tetrad. Its inverse field is
denoted by ${h^a}_\mu$ and satisfies the relations
\begin{eqnarray}
{h^a}_\mu{h_a}^\nu={\delta_\mu}^\nu, \quad
{h^a}_\mu{h_b}^\mu={\delta^a}_b.
\end{eqnarray}
In this paper, the Latin alphabet $(a,b,c,...=0,1,2,3)$ will be
used to denote the tangent space indices and the Greek alphabet
$(\mu,\nu,\rho,...=0,1,2,3)$ to denote the spacetime indices. The
Riemannian metric in TP theory arises as a by product [25] of the
tetrad field given by
\begin{equation}
g_{\mu\nu}=\eta_{ab}{h^a}_\mu{h^b}_\nu,
\end{equation}
where $\eta_{ab}$ is the Minkowski metric, i.e.,
$\eta_{ab}=diag(+1,-1,-1,-1)$. For the Weitzenb$\ddot{o}$ck
spacetime, the torsion is defined as
\begin{equation}
{T^\theta}_{\mu\nu}={\Gamma^\theta}_{\nu\mu}-{\Gamma^\theta}_{\mu\nu}
\end{equation}
which is antisymmetric w.r.t. its last two indices. Due to the
requirement of absolute parallelism, the curvature of the
Weitzenb$\ddot{o}$ck connection vanishes identically. The
Weitzenb$\ddot{o}$ck connection also satisfies the relation
\begin{equation}
{{\Gamma^{0}}^\theta}_{\mu\nu}={\Gamma^\theta}_{\mu\nu}
-{K^{\theta}}_{\mu\nu},
\end{equation}
where
\begin{equation}
{K^\theta}_{\mu\nu}=\frac{1}{2}({{T_\mu}^\theta}_\nu+{{T_\nu}^
\theta}_\mu-{T^\theta}_{\mu\nu})
\end{equation}
is the {\bf contortion tensor} and ${{\Gamma^{0}}^\theta}_{\mu\nu}$
are the Christoffel symbols in GR.

\section{TP Version of the Lie Derivative and the Killing Equations}

For a scalar, the TP Lie derivative will act as the directional
derivative which is given as
\begin{equation}
{{\cal L}^T}_{\bf{\xi}} \phi=\xi^\rho \nabla_\rho \phi=\xi^\rho
\frac{\partial \phi}{\partial \xi^\rho},
\end{equation}
where ${{\cal L}^T}_\xi$ is used to denote the TP Lie derivative
along the vector field $\xi$. The TP Lie derivative of a covariant
tensor of rank $2$ along a vector field $\xi$ is defined through the
covariant derivatives of that tensor and vector field as
\begin{equation}
({{\cal L}^T}_\xi \textbf{A})_{\mu \nu}=\xi^\rho \nabla_\rho
A_{\mu \nu}+(\nabla_\mu \xi^\rho )A_{\rho \nu}+(\nabla_\nu
\xi^\rho )A_{\mu \rho},
\end{equation}
where $\nabla_\rho$ stands for the TP covariant derivative and is
defined [25] as
\begin{equation}
\nabla_\rho A_{\mu\nu}=A_{\mu\nu,\rho}-
\Gamma^\theta_{\rho\nu}A_{\mu\theta}-
\Gamma^\theta_{\mu\rho}A_{\nu\theta},
\end{equation}
$\Gamma^\theta_{\rho\nu}$ are the Weitzenb$\ddot{o}$ck connection as
given by Eq.(1). In view of Eq.(9), Eq.(8) takes the form
\begin{eqnarray}
({{\cal L}^T}_\xi A)_{\mu \nu}&=&\xi^\rho( A_{\mu\nu,\rho} -
{\Gamma^\theta}_{\rho\nu}A_{\mu\theta}-
{\Gamma^\theta}_{\mu\rho}A_{\nu\theta})+A_{\rho
\nu}({\xi^\rho}_{,\mu}+{\Gamma^\rho}_{\theta\mu}\xi^\theta)\nonumber\\
&+&A_{\mu\rho}({\xi^\rho}_{,\nu}+{\Gamma^\rho}_{\theta\nu}\xi^\theta).
\end{eqnarray}
After some simple calculations and using Eq.(4), we get
\begin{eqnarray}
({{\cal L}^T}_\xi A)_{\mu \nu}= A_{\mu\nu,\rho}\xi^\rho +A_{\rho
\nu}{\xi^\rho}_{,\mu} + A_{\mu\rho}
{\xi^\rho}_{,\nu}+\xi^\rho(A_{\theta\nu}
{T^\theta}_{\mu\rho}+A_{\mu\theta} {T^\theta}_{\nu\rho}).
\end{eqnarray}
Similarly, the TP Lie derivative of a contravariant tensor of rank
$2$ can be written as
\begin{eqnarray}
({{\cal L}^T}_\xi A)^{\mu \nu}={ A^{\mu\nu}}_{,\rho}\xi^\rho
-A^{\rho \nu}{\xi^\mu}_{,\rho} + A^{\mu\rho}
{\xi^\nu}_{,\rho}-\xi^\rho(A^{\theta\nu}
{T^\mu}_{\theta\rho}+A^{\mu\theta} {T^\nu}_{\theta\rho}).
\end{eqnarray}
Following the same procedure, we can extend this definition for a
mixed tensor of rank $p+q$ as
\begin{eqnarray}
{({{\cal L}^T}_\xi A)^{\rho ... \sigma}}_{\mu...\nu}&=&\xi^\alpha
{A^{\rho ... \sigma}}_{\mu...\nu,\alpha} +{A^{\rho ...
\sigma}}_{\alpha...\nu} \xi^\alpha_{,\mu} + ... + {A^{\rho ...
\sigma}}_{\mu...\alpha} \xi^\alpha_{,\nu}\nonumber\\
&-&{A^{\alpha ... \sigma}}_{\mu...\nu} \xi^\rho_{,\alpha}- ...
-{A^{\rho ... \alpha}}_{\mu...\nu}
\xi^\sigma_{,\alpha}\nonumber\\
&+&\xi^\alpha({A^{\rho ... \sigma}}_{\beta...\nu}
{T^\beta}_{\mu\alpha}+...+{A^{\rho ... \sigma}}_{\mu...\beta}
{T^\beta}_{\nu\alpha}\nonumber\\
&-&{A^{\beta ... \sigma}}_{\mu...\nu}
{T^\rho}_{\beta\alpha}-...-{A^{\rho... \beta}}_{\mu...\nu}
{T^\sigma}_{\beta\alpha}),
\end{eqnarray}
where $|\{\rho ... \sigma\}|=p$ and $|\{\mu...\nu\}|=q$. Now, we can
define the TP Killing equations as
\begin{equation}
({{\cal L}^T}_{\bf{\xi}} g)_{\mu\nu}=0.
\end{equation}
Using Eq.(11), Eq.(14) becomes
\begin{eqnarray}
({{\cal L}^T}_\xi g)_{\mu \nu}= g_{\mu\nu,\rho}\xi^\rho +g_{\rho
\nu}{\xi^\rho}_{,\mu} + g_{\mu\rho}
{\xi^\rho}_{,\nu}+\xi^\rho(g_{\theta\nu}
{T^\theta}_{\mu\rho}+g_{\mu\theta} {T^\theta}_{\nu\rho}).
\end{eqnarray}

\section{TP Killing Vectors of the Einstein Universe}

The metric representing the Einstein universe is given as follows
\begin{equation}
ds^2=dt^2-\frac{1}{A^2(r)} dr^2-d\theta^2-r^2 sin^2\theta d\phi^2,
\end{equation}
where $A(r)=\sqrt{1-\frac{r^2}{R^2}}$ and $R$ is constant. Using
the procedure adopted in the papers [30-33], the tetrad components
of the above metric can be written as
\begin{equation}
{h^a}_\mu=\left\lbrack\matrix { 1   &&&   0    &&&   0    &&&   0
\cr 0        &&& \frac{1}{A}\sin\theta \cos\phi &&& r
\cos\theta\cos\phi &&& -r \sin\theta\sin\phi \cr 0        &&&
\frac{1}{A}\sin\theta \sin\phi &&& r \cos\theta\sin\phi &&& r
\sin\theta\cos\phi \cr 0        &&& \frac{1}{A}\cos\theta &&& -r
\sin\theta &&& 0\cr } \right\rbrack
\end{equation}
with its inverse
\begin{equation}
{h^a}_\mu=\left\lbrack\matrix { 1   &&&   0    &&&   0    &&&   0
\cr 0        &&& A \sin\theta \cos\phi &&& \frac{1}{r}
\cos\theta\cos\phi &&& -\frac{\sin\phi}{r \sin\theta} \cr 0 &&& A
\sin\theta \sin\phi &&& \frac{1}{r} \cos\theta\sin\phi &&&
\frac{\cos\phi}{r \sin\theta} \cr 0 &&& A \cos\theta &&&
-\frac{1}{r} \sin\theta &&& 0\cr } \right\rbrack.
\end{equation}
When we make use of Eqs.(17) and (18) in Eq.(1), we obtain the
following non-vanishing components of the Weitzenb$\ddot{o}$ck
connection
\begin{eqnarray}
{\Gamma^1}_{11}&=&-\frac{A'}{A}, \quad {\Gamma^1}_{22}=-rA,\quad {\Gamma^1}_{33}
=-rA \sin^2\theta,\nonumber\\
{\Gamma^2}_{12}&=&\frac{1}{rA}= {\Gamma^3}_{13},\quad
{\Gamma^2}_{21}=\frac{1}{r}= {\Gamma^3}_{31},\nonumber\\
{\Gamma^2}_{33}&=&-\sin\theta\cos\theta,\quad
{\Gamma^3}_{23}=\cot\theta={\Gamma^3}_{32}.
\end{eqnarray}
The corresponding non-vanishing components of the torsion tensor
are
\begin{eqnarray}
{T^2}_{12}=\frac{1}{r}(1-\frac{1}{A})={T^3}_{13}.
\end{eqnarray}
Making use of Eq.(20) in (15), we get the TP Killing equations in
the expanded from as
\begin{eqnarray}
\xi^1=A B(t,\theta,\phi),\\
\xi^1+rA \xi^2_{,2}=0,\\
\frac{1}{A}\xi^1 + r \cot\theta \xi^2 + r \xi^3_{,3}=0,\\
\xi^1_{,0}-A^2 \xi^0_{,1}=0,\\
\xi^0_{,2}-r^2 \xi^2_{,0}=0,\\
\xi^0_{,3}-r^2\sin^2\theta \xi^3_{,0}=0,\\
\xi^2_{,1}+\frac{1}{A^2r^2}\xi^1_{,2}+\frac{1}{r}(1-\frac{1}{A})\xi^2=0,\\
\xi^3_{,1}+\frac{1}{A^2r^2\sin^2\theta}\xi^1_{,3}+\frac{1}{r}(1-\frac{1}{A})\xi^3=0,\\
\xi^2_{,3}+ \sin^2 \theta \xi^3_{,2}=0,\\
\xi^0=N(r,\theta,\phi).
\end{eqnarray}
Solving Eqs.(21), (22) and (27) simultaneously, it follows that
\begin{eqnarray}
\xi^1&=& A \{ B_1(t,\phi)\cos\theta + B_2(t,\phi)\sin\theta \},\\
\xi^2&=& -\frac{1}{r} \{ B_1(t,\phi)\sin\theta -
B_2(t,\phi)\cos\theta \}+F(r)C(t,\phi),
\end{eqnarray}
where $F(r)=\frac{1}{r^2}(R-\sqrt{R^2-r^2})$. Substituting the value
of $\xi^2$ in Eq.(29), we get
\begin{eqnarray}
\xi^3&=&\frac{1}{r} \{ B_{1,3}(t,\phi)\ln|\csc\theta-\cot\theta|
+B_{2,3}(t,\phi)\csc\theta \} \nonumber\\
&+&F(r)C_{,3}(t,\phi)\cot\theta + D(t,r,\phi),
\end{eqnarray}
Using these values in the remaining equations, it follows that
\begin{eqnarray}
\xi^0&=&c_0,\nonumber\\
\xi^1&=& A \{c_1 \cos\theta + (c_2\cos\phi+c_3 \sin\phi)\sin \theta \},\nonumber\\
\xi^2&=& -\frac{1}{r} \{c_1\sin\theta -(c_2\cos\phi+
c_3 \sin\phi)\cos\theta\}\nonumber\\
&+&F(r)(c_4 \cos\phi+c_5 \sin\phi),\nonumber \\
\xi^3&=& \frac{1}{r} \{(c_3\cos\phi
-c_2\sin\phi)\csc\theta\}+c_6 F(r)\nonumber\\
&-&F(r)\cot\theta(c_4 \sin\phi-c_5 \cos\phi).
\end{eqnarray}
This gives the following $7$ KVs of the Einstein universe in the
context of TPG
\begin{eqnarray}
\xi_{(1)}&=&\frac{\partial}{\partial t},\nonumber\\
\xi_{(2)}&=& F(r) \frac{\partial}{\partial \phi}\nonumber\\
\xi_{(3)}&=& F(r)(\cos\phi \frac{\partial}{\partial
\theta}-\cot\theta\sin\phi \frac{\partial}{\partial
\phi}),\nonumber\\
\xi_{(4)}&=& F(r)(\sin\phi \frac{\partial}{\partial
\theta}+\cot\theta\cos\phi \frac{\partial}{\partial
\phi}),\nonumber\\
\xi_{(5)}&=& A \cos\theta \frac{\partial}{\partial
r}-\frac{1}{r}\sin \theta
\frac{\partial}{\partial \theta},\nonumber\\
\xi_{(6)}&=& A \sin\theta \cos\phi \frac{\partial}{\partial
r}+\frac{1}{r}\cos\phi \cos\theta \frac{\partial}{\partial
\theta}-\frac{1}{r}\sin\phi \csc\theta
\frac{\partial}{\partial \phi},\nonumber\\
\xi_{(7)}&=& A \sin\theta \sin\phi \frac{\partial}{\partial
r}+\frac{1}{r}\sin\phi \cos\theta \frac{\partial}{\partial
\theta}+\frac{1}{r}\cos\phi \csc\theta \frac{\partial}{\partial
\phi}.
\end{eqnarray}

\section{Summary and Discussion}

In this paper, we have defined the Lie derivative on the space
with torsion. This definition is then applied to find the TP
Killing vectors of the Einstein universe. It is shown that there
arise $7$ TP KVs which are the same in numbers as found in the
context of GR [2-3]. The comparison shows that $\xi_{(1)}$ is the
same in both the theories while $\xi_{(2)}$, $\xi_{(3)}$ and
$\xi_{(4)}$ in TPG are multiple of the corresponding KVs in GR by
$F(r)$.  For $F(r)=1$, that is, for $r=\pm \sqrt{2R-1}$, the first
four KVs reduce to the basic four KVs of the spherical symmetry
and coincide with those in GR. This implies that, in torsion
space, the spherical symmetry can be recovered for a particular
choice of $r=\pm \sqrt{2R-1}$. The remaining three TP KVs are
different from the KVs found in the context of GR. This difference
occurs due to the non-vanishing components of the torsion tensor
which involves in Eq.(15) when $\mu=\nu$. The extension of this
work to the other spacetimes is under investigation which may help
to make a conjecture about the relationship between the KVs in GR
and TP.

\vspace{0.5cm}

{\bf Acknowledgment}

\vspace{0.5cm}

We appreciate the Higher Education Commission Islamabad, Pakistan
for its financial support through the {\it Indigenous PhD 5000
Fellowship Program Batch-I} during this work.

\vspace{1.5cm}

{\bf References}

\begin{description}

\item{[1]} Petrov, A.Z.: Phys. {\it Einstein Spaces} (Pergamon, Oxford University Press, 1989).

\item{[2]} Bokhari, A.H. and Qadir, A.: J. Math. Phys. {\bf 31}(1990)1463.

\item{[3]} Bokhari, A.H. and Qadir, A.: J. Math. Phys. {\bf 34}(1993)3543.

\item{[4]} Ziad, M., Ph.D. Thesis (Quaid-i-Azam University, 1990).

\item{[5]} Qadir, A. and Ziad, M.: Nuovo Cimento {\bf B110}(1995)317.

\item{[6]} Stephani, H., Kramer, D., MacCallum, M., Hoenselaers and Herlt, E.:
\textit{Exact Solutions of Einstein's Field Equations} (Cambridge
University, 2003).

\item{[7]} Katzin, G.H., Levine, J. and Davis, H.R.: J. Math. Phys. {\bf
10}(1969)617.

\item{[8]} Katzin, G.H., Levine, J.: Colloquium Mathematicum(Poland) {\bf26}(1972)21.

\item{[9]} Hall, G.S. and da Costa, J.: J. Math. Phys. {\bf 32}(1991)2848.

\item{[10]} Hall, G.S. and da Costa, J.: J. Math. Phys. {\bf 32}(1991)2854.

\item{[11]} Amir, M.J., Bokhari, A.H. and Qadir, A.: J. Math. Phys. {\bf
35}(1994)3005.

\item{[12]} Bokhari, A.H. and Kashif, A.R.: J. Math. Phys. {\bf 37}(1996)3498.

\item{[13]} Feroze, T., Qadir, A. and Ziad, M.: J. Math. Phys. {\bf 41}(2000)2167.

\item{[14]} Carot, J., da Costa, J. and Vaz, E.G.L.R.: J. Math. Phys. {\bf 35}(1994)4852.

\item{[15]} Camci, U. and Sharif, M.: Class. Quantum Grav. {\bf 19}(2002)2169.

\item{[16]} Sharif, M. and Aziz, S.: Gen. Rel. Grav. {\bf 35}(2003)1093.

\item{[17]} Sharif, M.: J. Math. Phys. {\bf 44}(2003)5141.

\item{[18]} Sharif, M.: J. Math. Phys. {\bf 45}(2004)1518.

\item{[19]} Sharif, M.: J. Math. Phys. {\bf 45}(2004)1532.

\item{[20]} Hayashi, K. and Shirafuji, T.: Phys. Rev. {\bf D19}(1979)3524.

\item{[21]} Weitzenb$\ddot{o}$ck, R.: {\it Invarianten Theorie}(Gronningen: Noordhoft, 1923).

\item{[22]} De Andrade, V.C. and Pereira,  J.G.: Phys. Rev. {\bf
D56}(1997)4689.

\item{[23]} De Andrade, V.C. and Pereira,  J.G.: Gen. Rel. Grav. {\bf 30}(1998)263.

\item{[24]} Aldrovandi and Pereira, J.G.: {\it An Introduction to Geometrical Physics} (World Scientific, 1995).

\item{[25]} Aldrovendi, R. and Pereira, J.G.: {\it An Introduction to Gravitation Theory} (preprint).

\item{[26]} Hehl, F.W., McCrea, J.D., Mielke, E.W. and Ne'emann, Y.: Phys.
Rep. {\bf 258}(1995)1.

\item{[27]} Hammond, R.T.: Rep. Prog. Phys. {\bf 65}(2002)599.

\item{[28]} Nashed, G.G.L.: Phys. Rev. \textbf{D66}(2002)060415.

\item{[29]} Nashed, G.G.L.: Gen. Rel. Grav. \textbf{34}(2002)1074.

\item{[30]} Pereira, J.G., Vargas, T. and Zhang, C.M.: Class. Quantum Grav. {\bf 18}(2001)833.

\item{[31]} Sharif, M. and Amir, M.J.: Gen. Rel. Grav. {\bf 38}(2006)1735.

\item{[32]} Sharif, M. and Amir, M.J.: Gen. Rel. Grav. \textbf{39} (2007)989.

\item{[33]} Sharif, M. and Amir, M.J.: Mod. Phys. Lett. \textbf{A22}(2007)425.

\end{description}
\end{document}